\documentclass[aps,pra,twocolumn,reprint]{revtex4-1}
\usepackage{blindtext}

\usepackage{graphicx}
\usepackage{amssymb}
\usepackage{amsmath}
\usepackage{bm}
\usepackage{braket}
\usepackage{gensymb}
\usepackage{footmisc}
\usepackage{booktabs}
\usepackage{array}
\usepackage{hyperref}
\usepackage{xcolor}
\hypersetup{
    colorlinks,
    linkcolor={black},
    citecolor={blue},
    urlcolor={blue}
}

\begin{document}
\title{Photoassociation spectroscopy of ultracold $^{173}$Yb atoms near the intercombination line}
\author{Jeong Ho Han, Jin Hyoun Kang, Moosong Lee, Y. Shin}
\email{yishin@snu.ac.kr}

\affiliation{
Department of Physics and Astronomy and Institute of Applied Physics, Seoul National University, Seoul 08826, Korea
\\
Center for Correlated Electron Systems, Institute for Basic Science, Seoul 08826, Korea
}

\date{\today}

\begin{abstract}
We report on photoassociation (PA) spectroscopy of a degenerate Fermi gas of $^{173}$Yb atoms near the dissociation limit of the spin-forbidden ${^1}S_0$--${^3}P_1$ intercombination transition. An atom-loss spectrum is measured from a trapped sample for a spectral range down to $-1$~GHz with respect to the $f=5/2 \rightarrow f^\prime=7/2$ atomic resonance. The spectrum shows eighty PA resonances, revealing the high nuclear spin nature of the system. We investigate the Zeeman effect on the spectrum near a detuning of $-0.8$~GHz, where we examine the quantum numbers of the Zeeman levels using various two-component spin mixture samples. Finally, we measure the atom loss rate under PA light for several pronounced PA resonances. 
\end{abstract}

\maketitle

\section{Introduction}

Photoassociation (PA) is a process in which two colliding atoms form an excited molecule by absorbing a photon. PA spectroscopy provides a versatile tool for probing the physics of rovibrational molecular states~\cite{Jones_RMP_2006} and precisely determining the collisional properties of atoms, such as scattering length and interatomic potential coefficients~\cite{Kitagawa_Twocolor_2008}. Furthermore, the PA process can be actively used to control the strength of atomic interaction via coupling to an excited molecular state, a phenomenon called optical Feshbach resonance (OFR)~\cite{Fatemi_OFR_2000, Theis_OFR_2004, Ciurylo_OFR_2005, Enomoto_OFR_2008,Yamazaki2010,Yan2013}, and to measure pair correlations in strongly correlated atomic gas systems~\cite{Partridge_BCS_2005,Sugawa_dualMott_2011}.

Recently, there has been broad interest in studies on the PA physics of two-valence-electron atoms such as Yb~\cite{Takasu_PA1P1_2004,Tojo_PA_2006,Enomoto_OFR_2008,Enomoto_C6_2007, Kitagawa_Twocolor_2008,Borkowski_Lineshape_2009,Enomoto_PLR_2008,Borkowski_Hetero_2011, Roy_YbLi_2016}, Sr~\cite{Nagel_PASr_2005, Zelevinsky_Narrow_2006, Borkowski_Mass_2014, Stellmer_Sr2_2012, Reinaudi_Sr2_2012,Nicholson2015,Yan2013}, and Ca~\cite{Kahmann_Zeeman_2014,Tiemann_Zeeman_2015}. These atoms have narrow ${^1}S$--${^3}P$ intercombination transition, which is beneficial for the precise determination of PA resonances and enables the implementation of OFR without significant atom loss~\cite{Ciurylo_OFR_2005,Enomoto_OFR_2008,Yamazaki2010,Yan2013}. In particular, Yb atoms have rich, stable isotopes, including five spinless bosons ($^{168}$Yb, $^{170}$Yb, $^{172}$Yb, $^{174}$Yb, and $^{176}$Yb) and two fermions ($^{171}$Yb, with a nuclear spin of $i=1/2$ and $^{173}$Yb, with $i=5/2$), providing an interesting opportunity to study the mass scaling of PA physics~\cite{Kitagawa_Twocolor_2008,Borkowski_Lineshape_2009}. To date, many PA spectra of Yb atoms have been reported for bosonic isotopes~\cite{Tojo_PA_2006, Enomoto_C6_2007, Kitagawa_Twocolor_2008,Borkowski_Lineshape_2009}, fermionic $^{171}$Yb~\cite{Enomoto_PLR_2008}, and isotopic mixtures~\cite{Borkowski_Hetero_2011, Roy_YbLi_2016}. However, the complete PA spectrum of fermionic $^{173}$Yb, with its high nuclear spin, is still unknown, although a couple of PA resonances have been reported~\cite{Sugawa_dualMott_2011,Taie_Pomeranchuk_2012,Kitagawa_Twocolor_2008}. In addition, the $^{173}$Yb Fermi gas system has been discussed as a candidate platform for studies of exotic $SU(\mathcal{N}>2)$ quantum magnetism~\cite{Gorshkov_SUN_2010}. Information on the PA spectrum of $^{173}$Yb near the intercombination line is highly desirable for such a quantum simulation application~\cite{Reichenbach_OFR_2009}.

In this paper, we report the PA spectrum of a degenerate Fermi gas of $^{173}$Yb atoms near the dissociation limit of the $\ket{{^1}S_0,f=5/2}\rightarrow \ket{{^3}P_1,f'=7/2}$ intercombination transition. We measured an atom-loss spectrum as a function of the frequency of the PA light and we observed eighty PA resonances in the spectrum on the red-detuned side of the atomic resonance down to $-1$~GHz. The high density of the spectral lines can be attributed to the high nuclear spin number of $^{173}$Yb, which we confirmed by performing a multi-channel calculation of the molecular energy levels based on known spectroscopic results. To collect further spectroscopic information on the excited molecular states, we investigated the Zeeman effect in the spectrum near the frequency detuning of $-0.8$~GHz. By employing various two-component spin mixture samples, we determined the quantum numbers of the Zeeman sublevels and estimated the $g$ factor of the molecular state corresponding to the PA line at $-796$~MHz detuning. Finally, we measured the two-body loss rates under PA light for several pronounced PA resonances. Our measurement results provide a starting point for studies of the PA physics of fermionic $^{173}$Yb atoms, although further theoretical efforts will be required to interpret the measured spectra.

\section{Experiment}

\begin{center}
\begin{figure*}
\includegraphics[width=17.5cm]{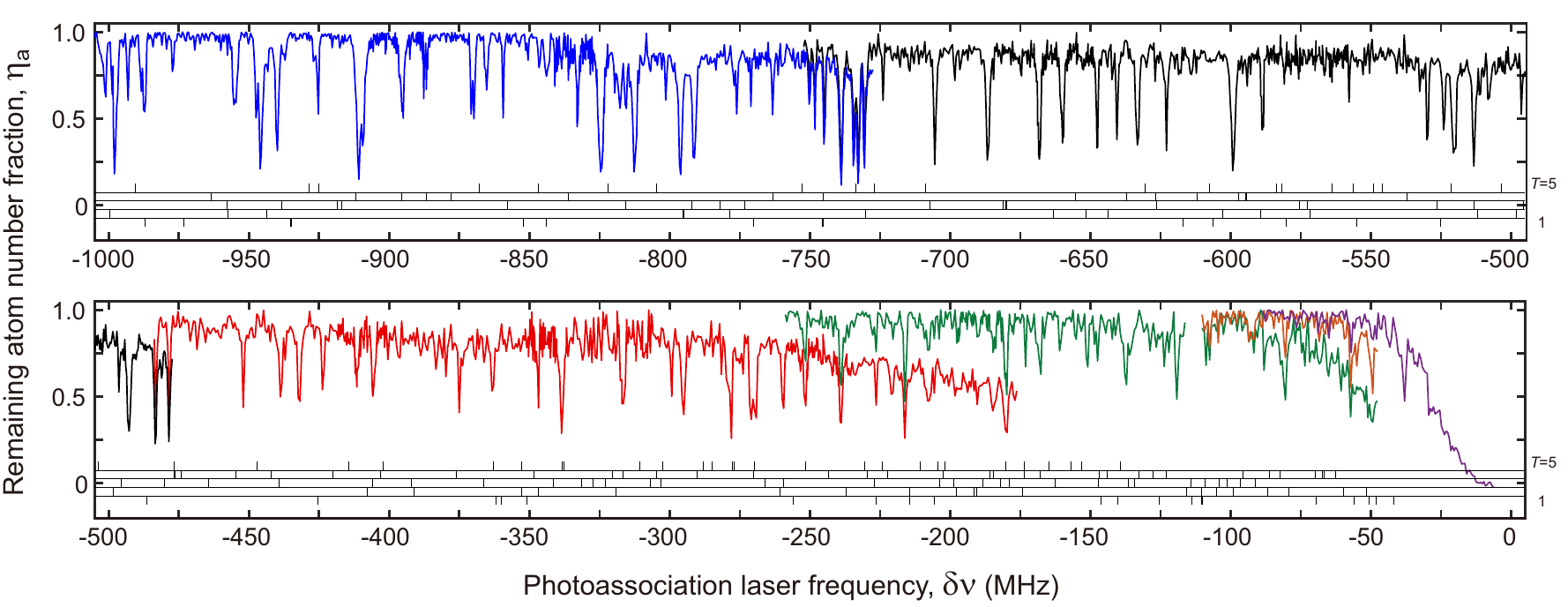}
\caption{ Photoassociation (PA) spectrum of unpolarized $^{173}$Yb atoms at low temperature. The various colors represent different PA laser beam conditions with intentisy $I_\mathrm{PA}$ and pulse duration $\tau$: blue ($I_\mathrm{PA}=$0.74~W/cm$^2$, $\tau=100$~ms), black (0.37~W/cm$^2$, $100$~ms), red (74~mW/cm$^2$, $100$~ms), green (74~mW/cm$^2$, $50$~ms), orange (74~mW/cm$^2$, $30$~ms), and purple (37~mW/cm$^2$, $30$~ms). The barcode lines at the bottom of the plot show the numerical calculation results obtained for the adiabatic molecular potentials with $T = 1$(bottom) to $5$(top) (Fig.~2).\label{fig:spectrum}}
\end{figure*}
\end{center}

We prepared an ultracold gas sample of $^{173}$Yb atoms as described in Ref.~\cite{Lee_SOC_2017}. The atoms were first collected in a magneto-optical trap using the 556~nm $^1S_0$--${^3}P_1$ intercombination transition and transferred into a 1070 nm optical dipole trap (ODT). Then, the atoms were transported to an auxiliary vacuum chamber by moving the focus of the trapping laser beam and were loaded into a crossed ODT formed by superposing an additional 532 nm trapping laser beam. The atomic sample was evaporatively cooled by lowering the trap depth, and after cooling, it was held for an additional 0.3~s to ensure equilibrium. The final sample was an equal mixture of all six spin components of the ${^1}S_0$ ground state, containing approximately $3.1\times 10^5$ atoms. The sample temperature was measured to be $T \approx 130$~nK. The \textit{in situ} density distribution of the trapped sample was found to be well fit by a Gaussian profile with a $1/e^2$ radius of $\{ \sigma_x, \sigma_y, \sigma_z\}\approx \{12.0, 7.5, 3.8\}~\mu$m, and the central density was estimated to be $n_0\approx 1.6\times 10^{14}$~cm$^{-3}$, corresponding to a Fermi energy of $E_F = \hbar^2 (\pi^2 n_0)^{2/3} /(2m) \approx k_B\times 190$~nK, where $\hbar$ is the Planck constant $h$ divided by $2\pi$, $m$ is the atomic mass, and $k_B$ is the Boltzmann constant. To measure the sample condition, we take absorption image using the ${^1}S_0$--${^1}P_1$ transition.

PA resonances were detected via atom loss by illuminating the trapped sample with a pulsed PA laser beam. The linewidth of our PA laser was $<70$~kHz, which is sufficiently narrow to probe excited molecular states with a natural linewidth of $\Gamma_\mathrm{nat}/2\pi \approx (2\Gamma_\mathrm{a})/2\pi = 364$~kHz, where $\Gamma_a$ is the atomic linewidth of the $^1S_0$--$^3P_1$ transition~\cite{Ciurylo_OFR_2005}. The PA laser beam was $\sigma^-$--polarized and focused onto the sample with a Gaussian beam waist of $\approx 114~\mu$m, which was large to uniformly irradiate the entire sample. We obtained a PA spectrum by measuring the remaining atom number fraction $\eta_a$ as a function of the frequency $\nu$ of the PA laser beam. For each $\nu$, we determined $\eta_a$ by measuring the numbers of atoms with and without application of the PA laser beam, respectively.

\section{Results}

\subsection{PA spectrum}

Figure~\ref{fig:spectrum} shows the PA spectrum measured for $\delta \nu=\nu-\nu_0= -1 \sim 0$~GHz, where $\nu_0$ is the resonance frequency for the $\ket{^1S_0,f=5/2}\rightarrow \ket{{^3}P_1,f^\prime=7/2}$ atomic transition. In the measurement, we reduced the PA beam intensity $I_\mathrm{PA}$ and the pulse duration $\tau$ in a piecewise manner as we approach the atomic resonance to avoid power broadening and photon scattering loss effects, where $I_\mathrm{PA}=0.037-0.74$~W/cm$^2$ and $\tau=30-100$~ms. The saturation intensity for the atomic transition is $I_\mathrm{sat}=0.14$~mW/cm$^2$. The spectrum shows a high density of spectral lines, and we identified eighty PA resonances in the range of $-1~\mathrm{GHz}<\delta \nu <-38$~MHz. For $\delta \nu >-38$~MHz, it was difficult to unambiguously identify PA resonances because of high photon scattering loss near the atomic resonance. The positions $\nu_b$ and linewidths $\Gamma_b$ of the spectral lines were determined from Lorentzian line fits to the measured data and are listed in Table~\ref{tab:spectrum}. In our experiment, the ac Stark shift due to the dipole trapping beams was $< 100$~kHz and insignificant, and the thermal broadening was negligible for $k_\mathrm{B} T/h \approx$ 4~kHz.  

\newcolumntype{M}[1]{>{\centering\arraybackslash}m{#1}}
\begin{centering}
\begin{table}[h]
  \centering
  \caption{Measured PA resonances $\nu_\mathrm{b}$ and the corresponding linewidths $\Gamma_\mathrm{b}$ from Lorentzian fits to the individual resonances depicted in Fig.~\ref{fig:spectrum}~\cite{Ciurylo_PACa_2004}. The errors represent the 95\% confidence intervals from the fits. For some resonances, no linewidth is given due to insufficient data points. Note that the PA laser beam intensity varies over the frequency detuning (see the caption of Fig.~1).}
  \label{tab:spectrum}

  \begin{tabular}{M{1.95cm}M{1.95cm}}
  \addlinespace
    \hline\hline
    $\nu_\mathrm{b}$  &  $\Gamma_\mathrm{b}/2\pi$  \\
    (MHz) & (MHz) \\
    \hline
	$-997.8\pm0.1$ & $1.3\pm0.2$ \\
	$-993.1\pm0.3$ & $0.6\pm0.4$ \\
	$-987.4\pm0.2$ & $1.6\pm0.6$ \\
	$-955.0\pm0.2$ & $1.5\pm0.5$ \\
	$-945.9\pm0.1$ & $2.1\pm0.4$ \\
	$-940.0\pm0.1$ & $1.2\pm0.3$ \\
	$-925.4\pm0.1$ & $0.5\pm0.4$ \\
	$-910.4\pm0.2$ & $2.8\pm0.5$ \\
	$-895.3\pm0.1$ & $1.3\pm0.3$ \\
	$-887.6\pm0.1$ & $0.5\pm0.2$ \\
	$-870.2\pm0.2$ & $1.6\pm0.5$ \\
	$-865.3\pm0.2$ & $0.9\pm0.4$ \\
	$-859.3\pm0.2$ & $0.4\pm0.3$ \\
	$-832.9\pm0.1$ & $0.8\pm0.3$ \\
	$-824.5\pm0.1$ & $2.1\pm0.3$ \\
	$-812.5\pm0.1$ & $1.5\pm0.4$ \\
	$-796.2\pm0.1$ & $1.4\pm0.2$ \\
	$-791.3\pm0.1$ & $1.3\pm0.2$ \\
	$-776.1\pm0.3$ & $0.6\pm0.4$ \\
	$-771.0\pm1.2$ & -- \\
	$-763.3\pm0.2$ & $0.6\pm0.3$ \\
	$-750.1\pm0.1$ &  --  \\
	$-748.3\pm0.1$ &  -- \\
	$-744.9\pm0.1$ & $0.5\pm0.3$ \\
	$-738.9\pm0.1$ & $1.1\pm0.2$ \\
	$-734.4\pm0.1$ & $0.7\pm0.3$ \\
	$-732.8\pm0.1$ & $0.9\pm0.2$ \\
	$-730.7\pm0.1$ & $0.5\pm0.2$ \\
	$-724.1\pm0.7$ & -- \\
	$-705.4\pm0.1$ & $0.8\pm0.2$ \\
	$-686.5\pm0.1$ & $1.0\pm0.3$ \\
	$-668.1\pm0.1$ & $1.0\pm0.2$ \\
	$-659.9\pm0.1$ & $1.1\pm0.3$ \\
	$-647.6\pm0.1$ & $0.5\pm0.3$ \\
	$-640.5\pm0.1$ & $0.5\pm0.3$ \\
	$-633.1\pm0.1$ & $1.0\pm0.3$ \\
	$-622.9\pm0.1$ & $0.6\pm0.2$ \\
	$-599.1\pm0.1$ & $1.4\pm0.3$ \\
	$-588.5\pm0.1$ & $0.7\pm0.3$ \\
	$-557.7\pm0.3$ & -- \\

    \hline\hline
  \end{tabular}
  \hfill
  \begin{tabular}{M{1.95cm}M{1.95cm}}
  \addlinespace
    \hline\hline
    $\nu_\mathrm{b}$  &  $\Gamma_\mathrm{b}/2\pi$  \\
    (MHz) & (MHz) \\
    \hline
	$-529.8\pm0.1$ & $0.6\pm0.5$ \\
	$-523.9\pm0.1$ & -- \\
	$-520.3\pm0.1$ & $1.5\pm0.4$ \\
	$-513.3\pm0.1$ & $1.1\pm0.3$ \\
	$-496.3\pm0.2$ & -- \\
	$-492.8\pm0.1$ & $1.3\pm0.4$ \\
	$-483.3\pm0.1$ & $1.1\pm0.3$ \\
	$-478.5\pm0.1$ & $0.9\pm0.3$ \\
	$-452.0\pm0.2$ & $0.6\pm0.3$ \\
	$-438.6\pm0.1$ & $1.2\pm0.5$ \\
	$-432.1\pm0.1$ & $0.8\pm0.5$ \\
	$-423.6\pm0.1$ & -- \\
	$-405.7\pm0.1$ & $1.1\pm0.6$ \\
	$-374.9\pm0.4$ & -- \\
	$-363.1\pm0.1$ & -- \\
	$-338.5\pm0.2$ & $1.3\pm0.5$ \\
	$-316.6\pm0.4$ & $1.1\pm0.3$ \\
	$-299.4\pm0.2$ & -- \\
	$-295.1\pm0.1$ & $1.3\pm0.4$ \\
	$-278.2\pm0.2$ & -- \\
	$-270.2\pm0.4$ & $2.3\pm1.3$ \\
	$-259.5\pm0.5$ & -- \\
	$-251.7\pm0.5$ & -- \\
	$-238.9\pm0.1$ & $0.9\pm0.7$ \\
	$-226.4\pm0.5$ & -- \\
	$-220.6\pm0.2$ & -- \\
	$-216.1\pm0.1$ & $0.7\pm0.3$ \\
	$-207.8\pm0.3$ & $1.9\pm0.9$ \\
	$-197.4\pm0.7$ & $2.6\pm2.2$ \\
	$-193.7\pm1.1$ & -- \\
	$-190.2\pm0.3$ & $1.1\pm0.9$ \\
	$-184.5\pm0.3$ & $1.9\pm0.9$ \\
	$-180.0\pm0.1$ & $1.5\pm0.4$ \\
	$-168.0\pm0.2$ & $1.4\pm0.9$ \\
	$-137.3\pm0.2$ & $1.5\pm0.8$ \\
	$-119.1\pm0.5$ & -- \\
	$-80.7\pm0.2$ & $1.0\pm0.7$ \\
	$-57.1\pm0.5$ & -- \\
	$-49.5\pm0.2$ & $1.2\pm0.9$ \\
	$-38.1\pm0.2$ & $0.9\pm0.7$ \\

    \hline\hline
  \end{tabular}
  
\end{table}
\end{centering}

\subsection{High spectral density}

To understand the observed high density of the spectral lines, we calculate the bound state energy levels for two $^{173}$Yb atoms in the $^3P_1$+$^1S_0$ channel, following the methods presented in Refs.~\cite{Zelevinsky_Narrow_2006,Reichenbach_OFR_2009}. The Hamiltonian for the two atoms is
\begin{equation} \label{eq:adiabeticpot} 
\hat{H}= \frac{p_r^2}{2\mu}+\frac{\hbar^2}{2\mu r^2} R(R+1) + V_\textrm{BO}(r) + \hat{H}_\mathrm{hf}, 
\end{equation}
where the first and second terms represent the radial and angular kinetic energies, respectively of the nuclei of the two atoms, $V_\textrm{BO}(r)$ is the electronic Born-Oppenheimer (BO) potential, and $\hat{H}_\mathrm{hf}$ is the hyperfine interaction term. Here, $\mu$ and $r$ denote the reduced mass and radial separation of the two nuclei, respectively, and $R$ is the quantum number for the overall rotation of the atom dimer. The BO potential is given by
\begin{equation} \label{eq:VBO}
V_\textrm{BO}(r)= -\frac{C_6}{r^6}\left(1-\frac{\sigma^6}{r^6}\right)-s\frac{C^{\Omega}_3}{r^3}.
\end{equation}
The first term is the Lennard-Jones potential and the second term represents the dipole-dipole interaction, where $s=+1 (-1)$ for the \textit{gerade} (\textit{ungerade}) potential and $\Omega$ is the internuclear projection of the angular momentum $\bm{J} = \bm{j}_1 + \bm{j}_2$. Here, $\bm{j}_{k=1,2}$ is the total electronic angular momentum of atom $k$. From Ref.~\cite{Borkowski_Lineshape_2009}, we have $C_6 = 2.41(0.22)\times 10^3 E_h a_0^6$, $\sigma =8.5(1.0)a_0$, and $C_3^0 = -2C_3^1 = -0.1949(11)\times E_h a_0^3$, where $E_h$ is the Hartree energy and $a_0$ is the Bohr radius. The hyperfine interaction is described by $\hat{H}_\mathrm{hf}=A(\bm{i}_1\cdot\bm{j}_1)+B\frac{3(\bm{i}_1\cdot\bm{j}_1)^2+\frac{3}{2}(\bm{i}_1\cdot\bm{j}_1)-i_1(i_1+1)j_1(j_1+1)}{2i_1j_1(2i_1-1)(2j_1-1)}$~\cite{Reichenbach_OFR_2009}, where we assume that the $k=2$ atom belongs to the ${^1}S_0$ state, i.e., $\bm{i}_2\cdot\bm{j}_2=0$. We adopt the values of $A/h =-1094.328$~MHz and $B/h = -826.635$~MHz from Ref.~\cite{Pandey_Hyperfine_2009}.

At the low temperature of our experiment, we expect only $s$-wave ($R=0$) collisions for two fermionic $^{173}$Yb atoms in the ${^1}S_0$ ground state, and the initial ${^1}S_0$+${^1}S_0$ dimer state should have total angular momentum of $T=F=I=0,2,$ or $4$ and even spatial parity ($p=1$). Here, $\bm{F} = \bm{f}_1 + \bm{f}_2$ and $\bm{I}=\bm{i}_1+\bm{i}_2$. According to the selection rules for optical excitation, excited molecular states should have $T=1,2,3,4,$ or $5$ and odd parity ($p=-1$).  In the modified Hund's case (e) that is relevant to our condition, with large spin-orbit coupling and hyperfine interaction, we count 205 different configurations of $(T,F,R)$ for the final states of the PA transition. Note that the transition from the initial $^1\Sigma_g$ molecular state to a \textit{gerade}-symmetry state is possible because the \textit{u-g} symmetry is broken in Hund's case (e)~\cite{Pique_ugsymmetry_1984}.

The adiabatic potentials for molecular states can be obtained by diagonalizing the Hamiltonian in Eq.~(1) via basis transformation between different Hund's cases~\cite{Tiesinga_PA_2005,Reichenbach_OFR_2009}. For a short distance $r$, the BO potential, which is diagonal under Hund's case (c), is dominant. In this case, the basis set is given by $\ket{\gamma} =\ket{J,\Omega,I,\iota,\Phi,(T,M_T,p)}$, where $\iota$ and $\Phi$ are the projections of $\bm{I}$ and $\bm{F}$ onto the internuclear axis, respectively, and $M_T$ is the projection of the total angular momentum onto a space-fixed quantization axis. At large $r$, i.e., when the two atoms are far apart, Hund's case (p) becomes relevant and results in a basis set consisting of the products of internal atomic states and molecular rotation as follows: $\ket{\pi}=\ket{f_1,m_1,f_2,m_2,R,(T,M_T,p)}$. In the intermediate range of $r$, we consider Hund's case (e), in which rotational and hyperfine interactions are diagonal and use a basis of $\ket{\epsilon}=\ket{f_1,f_2,F,R,(T,M_T,p)}$.

\begin{figure}
\includegraphics[width=8.0cm]{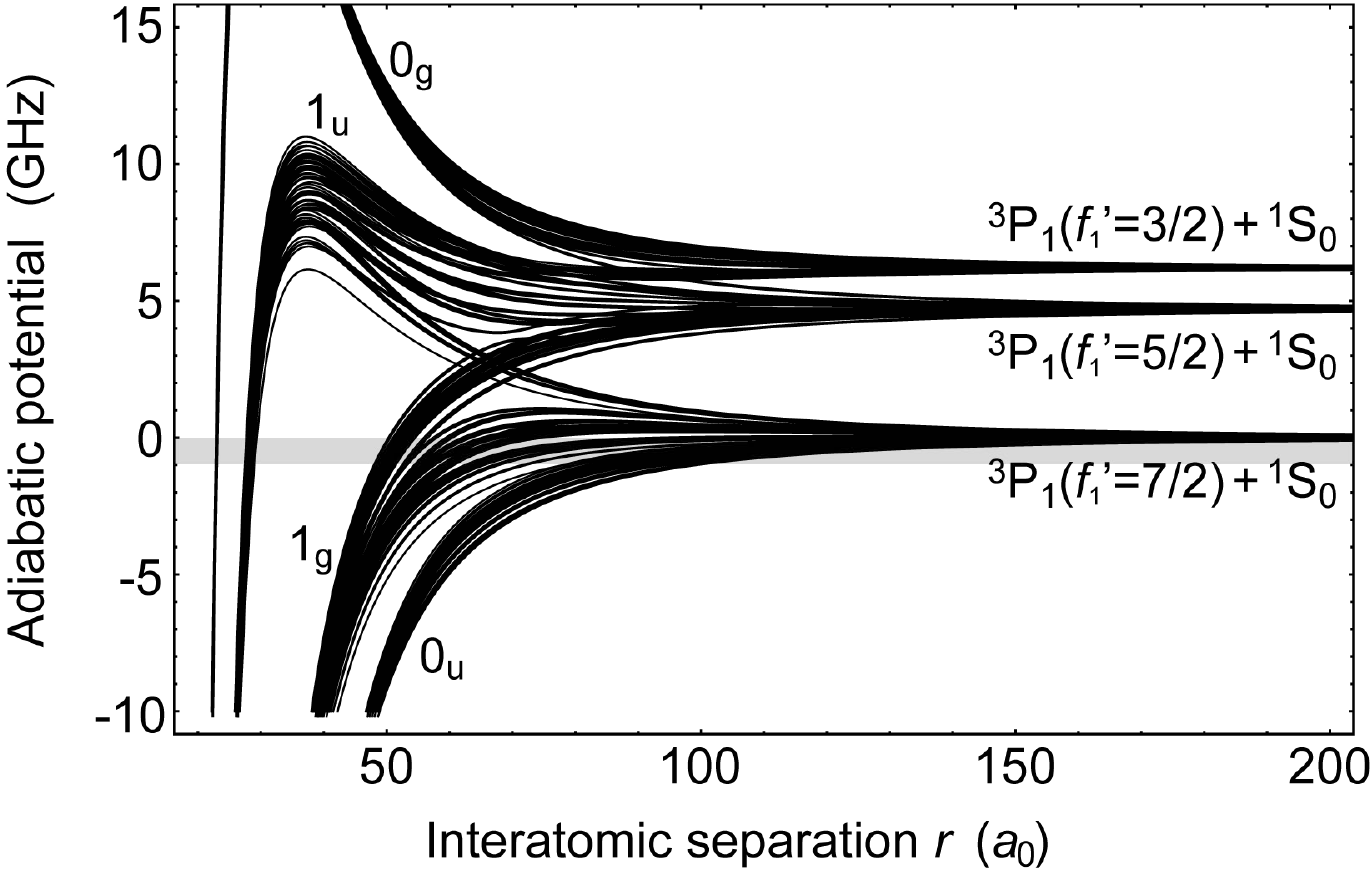}
\caption{Adiabatic molecular potentials for a $^{173}$Yb$_2$ dimer in the $^1S_0+$$^3P_1$ channel as functions of the interatomic separation $r$. The molecular potentials for 205 different $(T,F,R)$ configurations are displayed, which are accessible via PA from the initial $s$-wave colliding atoms in the $^1S_0+$$^1S_0$ channel. At large $r$, the potentials converge to three asymptotic branches, which correspond to excited atomic states with hyperfine numbers of $f_1^\prime=3/2, 5/2$, and $7/2$. The energy offset is adjusted to the $f_1^\prime=7/2$ asymptote. The shaded region indicates the spectral range of our measurements.\label{fig:adiabaticpot}
}
\end{figure}

The calculated adiabatic potentials for the 205 channels for excited molecular states are displayed in Fig.~\ref{fig:adiabaticpot}. At small $r$, the potentials are grouped into four branches, representing the four different dipole-dipole interaction configurations, and at large $r$, they converge to three asymptotes near the dissociation limit, which correspond to $f_1^\prime=3/2, 5/2,$ and $7/2$, respectively. We note that the potential related to the $f_1^\prime=3/2$ asymptote has a local minimum near $\sim 75~a_0$, predicting purely-long-range molecular states. This is due to the large hyperfine structure of heavy Yb atoms, and purely-long-range states have been observed with $^{171}$Yb~\cite{Enomoto_PLR_2008}.

From the calculated molecular potentials, we compute the bound state energy levels using a multi-channel discrete variable representation (DVR) method~\cite{Tiesinga_PA_2005,Tiesinga_DVR_1998,Reichenbach_OFR_2009}. This calculation predicts more than 200 bound states in the range of $-1~\mathrm{GHz} < \delta \nu <-38$~MHz, whose positions are indicated in Fig.~\ref{fig:spectrum} alongside the measured PA spectrum~\cite{LBE_comment}. Considering the limited experimental sensitivity, the observed high density of the spectral lines is reasonably explained by the calculated results. With regard to the resonance positions, a better comparison might be enabled by using an iterative fitting method to tune the potential coefficients values~\cite{Borkowski_Lineshape_2009}, but because of the heavy calculation load involved, we will leave such an effort as a topic for future studies.

\begin{figure}
\includegraphics[width=8.0cm]{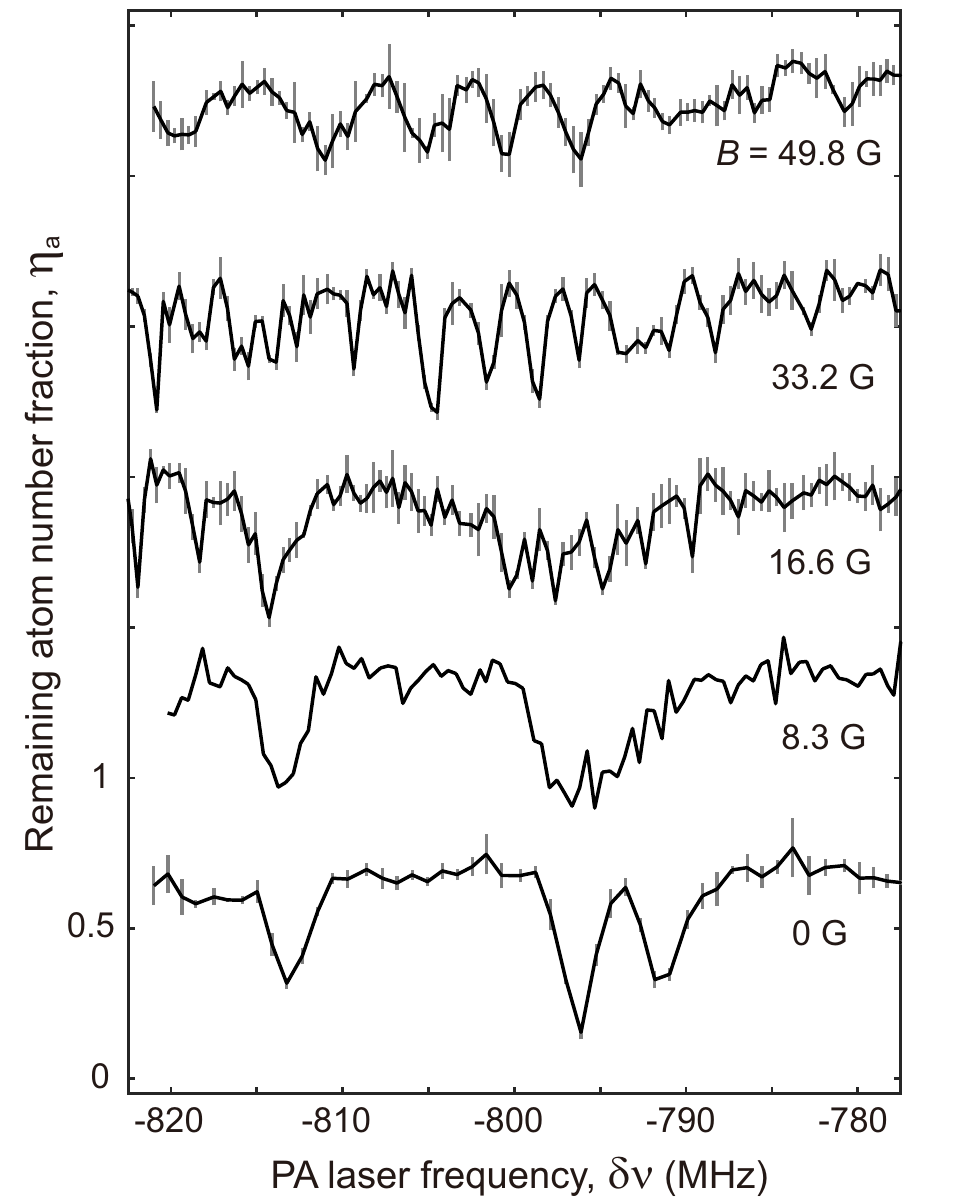}
\caption{PA spectra near $\delta \nu =-800$~MHz for various magnetic fields of $B=0$~G, $8.3$~G, $16.6$~G, $33.2$~G, and $49.8$~G. The PA laser beam was $\sigma^{-}$-polarized and the magnetic field was applied along the beam axis. All data points except those at $B=8.3$~G were obtained by averaging five independent measurements and the error bars denote their standard deviations. The data are offset for clarity.
\label{fig:Bchange}}
\end{figure}

\begin{figure}
\includegraphics[width=8.2cm]{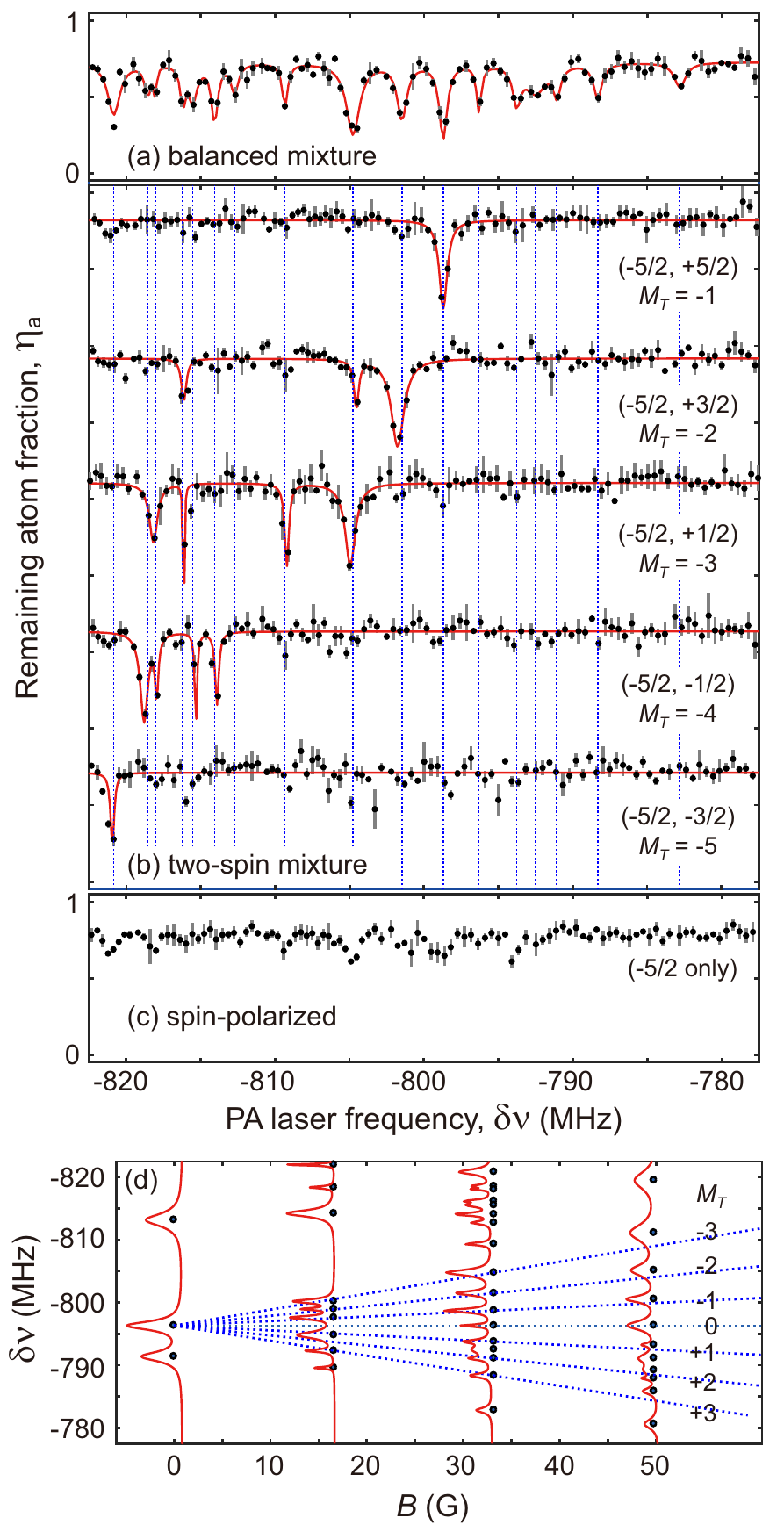}
\caption{PA spectra of various spin mixtures for $B=33.2$~G: (a) an unpolarized spin mixture, as shown in Fig.~\ref{fig:Bchange}; (b) two-component spin mixtures; and (c) a spin-polarized Fermi gas. The red lines show the sums of the Lorentzian fit curves to guide the eye, and the dotted vertical lines indicate the resonance positions as fitted from the spectrum of the balanced mixture. All data points were obtained by averaging five independent measurements of the same experiment and the error bars denote their standard deviations. (d) Zeeman splitting of the $-796$~MHz resonance (dotted lines), with markers representing the measured PA resonance positions.
\label{fig:Bchangecomp}}
\end{figure}

\subsection{Zeeman effect}

To facilitate the spectroscopic identification of the observed PA lines, we investigated the Zeeman effect by applying an external magnetic field $B$~\cite{Hamley_RARb_2009}. In the presence of $B$, the total angular momentum $T$ and its projection $M_T$ onto the field direction are still good quantum numbers of the system, and for low $B$ the Zeeman shift is described as $\Delta E_Z=\mu_B g B M_T$, where $\mu_B$ and $g$ are the Bohr magneton and the Lande $g$-factor of the molecular state, respectively. The number of Zeeman sublevels and the magnitude of their spectral splitting for $B$ directly reveal the quantum numbers $(T,M_T)$ of the molecular state as well as the $g$ factor value, which is sensitively determined by the interatomic potential~\cite{Kahmann_Zeeman_2014, Tiemann_Zeeman_2015}.

We applied the magnetic field along the axis of the PA laser beam and measured the PA spectra for various $B$ (Fig.~\ref{fig:Bchange}). The scan range of $\delta \nu$ was set to be from $-820~\mathrm{MHz}$ to $-780~\mathrm{MHz}$, where for $B=0$~G, three pronounced PA lines are located at $\delta \nu \approx-813$~MHz, $-796$~MHz, and $-791$~MHz, with relatively large linewidths of $>1$~MHz. The latter two PA lines have been reported in previous experiments~\cite{Kitagawa_Twocolor_2008,Taie_Pomeranchuk_2012,Sugawa_dualMott_2011}. With increasing $B$, each spectral peak broadens and splits into multiple weak peaks. The Zeeman splitting response appears relatively rapidly for the line at $-791$~MHz and slowly for the line at $-813$~MHz, reflecting the different magnitudes of $gT$ for these PA lines. Asymmetric shifting toward negative detuning is observed, which we attribute mainly to the $\sigma^{-}$ polarization of the PA light, which allows only $\Delta M_T = -1$ transitions.

For $B>30$~G, the spectrum shows a group of fully resolved Zeeman peaks with linewidths of $\sim 1$~MHz. To determine the $M_T$ numbers of the Zeeman peaks, we measured the PA spectra of two-component spin mixture samples. When such a sample is prepared with two spin components with magnetic Zeeman numbers of $m_{f_1}$ and $m_{f_2}$, the initial dimer state for $s$-wave collision in $^1S_0$+$^1S_0$ has a specific quantum number of $M_{T}=m_{f_1}+m_{f_2}$, and with a $\sigma^{-}$--polarized PA laser beam, this state can be coupled only to excited molecular states with $M_T=m_{f_1}+m_{f_2}-1$. Thus, the corresponding $M_T$ number can be assigned to Zeeman peaks that appear in the PA spectrum of such a two-component sample. In our experiment, we employed five binary spin mixtures of $m_{f_1}=-5/2$ and $m_{f_2}=\{-3/2,-1/2,1/2,3/2,5/2\}$~\cite{Lee_SOC_2017} and the PA spectra of the samples were measured at $B=33.2$~G [Fig.~\ref{fig:Bchangecomp}(b)]. As expected, each spectrum shows a subset of the Zeeman peaks observed in the PA spectrum of a fully balanced spin mixture [Fig.~\ref{fig:Bchangecomp}(a)]. To suppress unwanted optical pumping by the PA light into different spin states, we set $I_\mathrm{PA}=0.16$~W/cm$^2$ and $\tau = 100$~ms to obtain $\Gamma_\mathrm{sc}\tau \approx 0.8$, where $\Gamma_\mathrm{sc}$ is the Rayleigh scattering rate of the PA light at $\delta \nu = -800$~MHz.  

The main finding from the $M_T$ analysis is that the three Zeeman peaks that are almost equally spaced in the detuning range of $-805~\mathrm{MHz}<\delta \nu <-798~\mathrm{MHz}$ have $M_T=-3,-2,$ and $-1$, respectively. We find that Zeeman peaks are also located at the positions linearly extrapolated for $M_T=0,1,2,$ and $3$ from these three Zeeman peaks and, in particular, that the peak position corresponding to $M_T=0$ coincides with the zero-field PA line at $-796$~MHz. From these observations and the fact that there is no $M_T=-4$ Zeeman peak at the corresponding expected spacing from the $M_T=-3$ peak, we infer that the total angular momentum number of the PA line at $\delta \nu=-796.2$~MHz is $T=3$. From a linear fit to the seven Zeeman peak positions, a $g$-factor of $g=0.056(3)$ can be determined, which is approximately ten times smaller than the atomic value of $g_F=0.426$ for the $^3P_1$ state. In Fig.~\ref{fig:Bchangecomp}(d), we display the PA resonance positions measured from the data in Fig.~\ref{fig:Bchange} as a function of $B$, and the Zeeman splitting lines predicted from the measured $g$-factor are found to reasonably fit the experimental data. For a high $B$ of approximately $\approx 50$~G, the PA resonance positions  slightly deviate from the predictions toward a negative detuning except for $M_T=0$, indicating higher-order Zeeman effects.

Although the $M_T$ information is helpful for deciphering the linear Zeeman splitting of the PA line at $-796$~MHz, an analysis of the Zeeman effects of the other two PA lines is not straightforward. First, we observe no $M_T=-1$ Zeeman peaks for these two PA lines, although such peaks should exist because $T\geq 1$. Second, each PA spectrum for $M_T=-3$ and $-4$ shows four resonances [Fig.~4(b)], which means that our PA spectrum for a high $B$ of $>30$~G must involve Zeeman contributions from other PA lines outside the measurement range. Theoretical support will be critical for a complete understanding of the observed Zeeman effects.

\begin{figure}
\includegraphics[width=7.6cm]{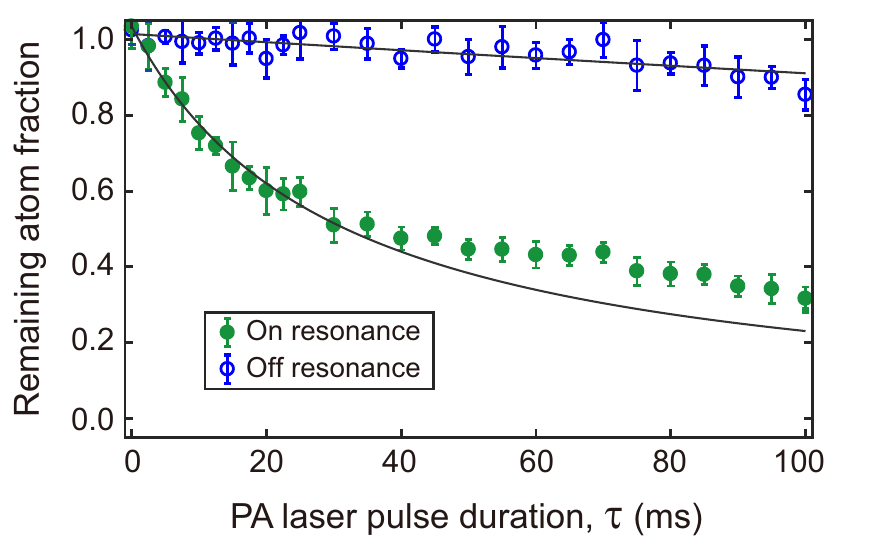}
\caption{Atom loss curves as functions of the pulse duration $\tau$ of the PA light for $\delta \nu =-791$~MHz (on resonance, solid green circles) and $\delta \nu =-784$~MHz (off resonance, open blue circles) with $I_\mathrm{PA}=$0.74~W/cm$^2$. The solid lines are exponential fits to the initial $10$~ms of decay data. All data points were obtained by averaging the results of five independent runs of the same experiment.
\label{fig:decay}}
\end{figure}

\subsection{Two-body loss rate}

Finally, we characterized some of the pronounced PA resonances by measuring the two-body loss rate $K_2$ under PA light. $K_2$ contains important information such as the Franck-Condon factor for the optical transition~\cite{Jones_RMP_2006,Bohn_semianalytic_1999} and the so-called optical length $l_\mathrm{opt}$ that describes the magnitude of the change of the scattering length due to the OFR~\cite{Ciurylo_OFR_2005,Borkowski_Lineshape_2009,Nicholson2015}. 

In the presence of PA light, the atom density $n$ evolves as $\dot{n}(t)=-2 K_2 n^2 - \gamma n$, where the first term represents the two-body PA process and the second term accounts for one-body decay processes such as Rayleigh photon scattering loss and background trap loss. For a case of a trapped sample, considering its inhomogeneous density distribution, the rate equation for the total number of atoms $N$ is given by  $\dot{N}(t) = - 2 K_2 \frac{N^2}{V_e} - \gamma N$, where $V_e=(2\pi)^{3/2} \sigma_x \sigma_y \sigma_z$ is the effective volume of the sample for a Gaussian density distribution. To avoid nonlinear effects caused by sample heating on $K_2$ and $V_e$, we measured the decay rate $\gamma'$ of $N$ from an exponential fit to the initial $10$~ms of $N(t)$ data and calculated $K_2$ as $K_2=\frac{V_e}{2\bar{N}}(\gamma'-\gamma)$. Here $\bar{N}$ denotes the average number of atoms over the initial 10~ms and $\gamma$ was independently measured at off-resonance detuning which is more than $4\Gamma_b$ away from the PA resonance (Fig.~\ref{fig:decay}). 

We measured $K_2$ for the three PA resonances at $\delta \nu=-38.1$~MHz, $-791.3$ ~MHz, and $-796.2$~MHz, and obtained $K_2=1.0(3)\times 10^{-12}$~cm$^{3}$/s with $I_\mathrm{PA}=74$~mW/cm$^{2}$, $K_2=0.5(1)\times 10^{-12}$~cm$^{3}$/s with $I_\mathrm{PA}=0.74$~W/cm$^{2}$, and $K_2=0.8(5)\times 10^{-12}$~cm$^{3}$/s with $I_\mathrm{PA}=0.74$~W/cm$^{2}$, respectively.  In the cold collision limit, the optical length is given as $l_\mathrm{opt}= \eta \mu K_2 /(8\pi \hbar)$~\cite{Ciurylo_OFR_2005,Yan2013}, where $\eta$ is the enhancement factor of the molecular linewidth with respect to the natural linewidth. Assuming that $\eta$ is order of unity, our measurement results suggest that $l_\mathrm{opt} \sim 10 a_0$ at $I_\mathrm{PA}=1$~W/cm$^{2}$. It is surprising that the estimated value of $l_\mathrm{opt}$ is more than two orders of magnitude smaller than the values reported for other Yb isotopes~\cite{Reichenbach_OFR_2009,Borkowski_Lineshape_2009,Kim_OFR_2016}. It would be worthwhile to investigate the tempearture and $\delta \nu$ dependence of $K_2$ in a further systematic manner~\cite{Borkowski_Lineshape_2009}.

\section{Summary}

We have measured the PA spectrum of a degenerate Fermi gas of $^{173}$Yb atoms near the dissociation limit of the ${^1}S_0$--${^3}P_1$ intercombination transition and have characterized some of the prominent PA lines by investigating their Zeeman splitting and measuring their two-body loss rates under PA light. The high density of the spectral lines was accounted for by the calculation of the molecular energy levels based on an extended version of Hund's case (e), but further theoretical investigation will be necessary for spectral identification of the observed molecular states. This will improve our understanding of the collisional properties of Yb atoms in $^1S$+$^3P$, which are important for many applications using Yb atoms, such as optical clocks~\cite{Bloom_clock_2014,Nemitz2016} and the simulation of novel quantum magnetism~\cite{Sugawa_dualMott_2011,Gorshkov_SUN_2010}.

Finally, we note that when the Zeeman splitting of a PA line is fully understood in terms of its molecular quantum number, the spin-dependent PA transitions may find immediate use in probing interspin correlations, particularly, in optical lattice experiments~\cite{Sugawa_dualMott_2011,Hofrichter2016}. For example, the correlations between the $m_f=m$ and $-m$ spin states may be distinctively detected by using the PA resonance at $-798.4$~MHz for $B=33.2$~G. Thus, it might be worthwhile to search for an isolated $T=5$ PA line whose Zeeman lines are spectroscopically well resolved and have reasonable transition strengths for a moderate magnetic field.

\section{Acknowledgments}
We thank Paul S. Julienne and Jee Woo Park for helpful discussions. This work was supported by the Institute for Basic Science (IBS-R009-D1) and the National Research Foundation of Korea (Grant No. 2014-H1A8A1021987).


\begin{references}

\bibitem{Jones_RMP_2006} K. M. Jones, E. Tiesinga, P. D. Lett, and P. S. Julienne, \href{https://doi.org/10.1103/RevModPhys.78.483}{Rev. Mod. Phys. $\bm{78}$, 483 (2006).}

\bibitem{Kitagawa_Twocolor_2008} M. Kitagawa, K. Enomoto, K. Kasa, Y. Takahashi, R. Ciury{\l}o, P. Naidon, and P. S. Julienne
, \href{https://doi.org/10.1103/PhysRevA.77.012719}{Phys. Rev. A $\bm{77}$, 012719 (2008)}.

\bibitem{Fatemi_OFR_2000} F. K. Fatemi, K. M. Jones, and P. D. Lett, \href{https://doi.org/10.1103/PhysRevLett.85.4462}{Phys. Rev. Lett. $\bm{85}$, 4462 (2000).}

\bibitem{Theis_OFR_2004} M. Theis, G. Thalhammer, K. Winkler, M. Hellwig, G. Ruff, R. Grimm, and J. Hecker Denschlag, \href{https://doi.org/10.1103/PhysRevLett.93.123001}{Phys. Rev. Lett. $\bm{93}$, 123001 (2004).}

\bibitem{Ciurylo_OFR_2005} R. Ciury{\l}o, E. Tiesinga, and P. S. Julienne, \href{https://doi.org/10.1103/PhysRevA.71.030701}{Phys. Rev. A $\bm{71}$, 030701(R) (2005).}

\bibitem{Enomoto_OFR_2008} K. Enomoto, K. Kasa, M. Kitagawa, and Y. Takahashi, \href{https://doi.org/10.1103/PhysRevLett.101.203201}{Phys. Rev. Lett. $\bm{101}$, 203201 (2008).}

\bibitem{Yamazaki2010} R. Yamazaki, S. Taie, S. Sugawa, and Y. Takahashi, \href{https://doi.org/10.1103/PhysRevLett.105.050405}{Phys. Rev. Lett. $\bm{105}$, 050405 (2010).}

\bibitem{Yan2013} M. Yan, B. J. DeSalvo, B. Ramachandhran, H. Pu, and T. C. Killian, \href{https://doi.org/10.1103/PhysRevLett.110.123201}{Phys. Rev. Let. $\bm{110}$, 123201 (2013)}.

\bibitem{Partridge_BCS_2005} G. B. Partridge, K. E. Strecker, R. I. Kamar, M. W. Jack, and R. G. Hulet, \href{https://doi.org/10.1103/PhysRevLett.95.020404}{Phys. Rev. Lett. $\bm{95}$, 020404 (2005).}

\bibitem{Sugawa_dualMott_2011} S. Sugawa, K. Inaba, S. Taie, R. Yamazaki, M. Yamashita, and Y. Takahashi, \href{https://doi.org/10.1038/nphys2028}{Nat. Phys. $\bm{7}$, 642 (2011).}

\bibitem{Takasu_PA1P1_2004} Y. Takasu, K. Komori, K. Honda, M. Kumakura, T. Yabuzaki, and Y. Takahashi, \href{https://doi.org/10.1103/PhysRevLett.93.123202}{Phys. Rev. Lett. $\bm{93}$, 123202 (2004).}

\bibitem{Tojo_PA_2006} S. Tojo, M. Kitagawa, K. Enomoto, Y. Kato, Y. Takasu, M. Kumakura, and Y. Takahashi, \href{https://doi.org/10.1103/PhysRevLett.96.153201}{Phys. Rev. Lett. $\bm{96}$, 153201 (2006).}

\bibitem{Enomoto_C6_2007} K. Enomoto, M. Kitagawa, K. Kasa, S. Tojo, and Y. Takahashi, \href{https://doi.org/10.1103/PhysRevLett.98.203201}{Phys. Rev. Lett. $\bm{98}$, 203201 (2007).}

\bibitem{Enomoto_PLR_2008} K. Enomoto, M. Kitagawa, S. Tojo, and Y. Takahashi, \href{https://doi.org/10.1103/PhysRevLett.100.123001}{Phys. Rev. Lett. $\bm{100}$, 123001 (2008).}

\bibitem{Borkowski_Lineshape_2009} M. Borkowski, R. Ciury{\l}o, P. S. Julienne, S. Tojo, K. Enomoto, and Y. Takahashi, \href{https://doi.org/10.1103/PhysRevA.80.012715}{Phys. Rev. A $\bm{80}$, 012715 (2009).}

\bibitem{Borkowski_Hetero_2011} M. Borkowski \textit{et al}., \href{https://doi.org/10.1103/PhysRevA.84.030702}{Phys. Rev. A $\bm{84}$, 030702(R) (2011).}

\bibitem{Roy_YbLi_2016} R. Roy, R. Shrestha, A. Green, S. Gupta, M. Li, S. Kotochigova, A. Petrov, and C. H. Yuen, \href{https://doi.org/10.1103/PhysRevA.94.033413}{Phys. Rev. A $\bm{94}$, 033413 (2016).}

\bibitem{Nagel_PASr_2005} S. B. Nagel, P. G. Mickelson, A. D. Saenz, Y. N. Martinez, Y. C. Chen, T. C. Killian, P. Pellegrini, and R. C{\^o}t{\'e}, \href{https://doi.org/10.1103/PhysRevLett.94.083004}{Phys. Rev. Lett. $\bm{94}$, 083004 (2005).}

\bibitem{Zelevinsky_Narrow_2006} T. Zelevinsky, M. M. Boyd, A. D. Ludlow, T. Ido, J. Ye, R. Ciury{\l}o, P. Naidon, and P. S. Julienne, \href{https://doi.org/10.1103/PhysRevLett.96.203201}{Phys. Rev. Lett $\bm{96}$, 203201 (2006).}

\bibitem{Stellmer_Sr2_2012} S. Stellmer, B. Pasquiou, R. Grimm, and F. Schreck, \href{https://doi.org/10.1103/PhysRevLett.109.115302}{Phys. Rev. Lett. $\bm{109}$, 115302 (2012).}

\bibitem{Reinaudi_Sr2_2012} G. Reinaudi, C. B. Osborn, M. McDonald, S. Kotochigova, and T. Zelevinsky, \href{https://doi.org/10.1103/PhysRevLett.109.115303}{Phys. Rev. Lett. $\bm{109}$, 115303 (2012).}

\bibitem{Borkowski_Mass_2014} M. Borkowski, P. Morzy{\'n}ski, R. Ciury{\l}o, P. S. Julienne, M. Yan, B. J. DeSalvo, and T. C. Killian, \href{https://doi.org/10.1103/PhysRevA.90.032713}{Phys. Rev. A $\bm{90}$, 032713 (2014).}

\bibitem{Nicholson2015} T. L. Nicholson, S. Blatt, B. J. Bloom, J. R. Williams, J. W. Thomsen, J. Ye, and P. S. Julienne, \href{https://doi.org/10.1103/PhysRevA.92.022709}{Phy. Rev. A $\bf{92}$, 022709 (2015).}

\bibitem{Kahmann_Zeeman_2014} M. Kahmann, E. Tiemann, O. Appel, U. Sterr, and F. Riehle, \href{https://doi.org/10.1103/PhysRevA.89.023413}{Phys. Rev. A $\bm{89}$, 023413 (2014).}

\bibitem{Tiemann_Zeeman_2015} E. Tiemann, M. Kahmann, E. Pachomow, F. Riehle, and U. Sterr, \href{https://doi.org/10.1103/PhysRevA.92.023419}{Phys. Rev. A $\bm{92}$, 023419 (2015).}

\bibitem{Taie_Pomeranchuk_2012} S. Taie, R. Yamazaki, S. Sugawa, and Y. Takahashi, \href{https://doi.org/10.1038/nphys2430}{Nat. Phys. $\bm{8}$, 825 (2012).}

\bibitem{Gorshkov_SUN_2010} A. V. Gorshkov, M. Hermele, V. Gurarie, C. Xu, P. S. Julienne, J. Ye, P. Zoller, E. Demler, M. D. Lukin, and A. M. Rey, \href{https://doi.org/10.1038/nphys1535}{Nat. Phys. $\bm{6}$, 289 (2010).}

\bibitem{Reichenbach_OFR_2009} I. Reichenbach, P. S. Julienne, and I. H. Deutsch, \href{https://doi.org/10.1103/PhysRevA.80.020701}{Phys. Rev. A $\bm{80}$, 020701(R) (2009).}

\bibitem{Lee_SOC_2017} M. Lee, J. H. Han, J. H. Kang, and Y. Shin, \href{https://doi.org/10.1103/PhysRevA.95.043627}{Phys. Rev. A $\bm{95}$, 043627 (2017).}

\bibitem{Ciurylo_PACa_2004} R. Ciury{\l}o, E. Tiesinga, S. Kotochigova, and P. S. Julienne, \href{https://doi.org/10.1103/PhysRevA.70.062710}{Phys. Rev. A $\bm{70}$, 062710 (2004).}

\bibitem{Pandey_Hyperfine_2009} K. Pandey, A. K. Singh, P. V. K. Kumar, M. V. Suryanarayana, and V. Natarajan, \href{https://doi.org/10.1103/PhysRevA.80.022518}{Phys. Rev. A $\bm{80}$, 022518 (2009).}

\bibitem{Pique_ugsymmetry_1984} J. P. Pique, F. Hartmann, R. Bacis, S. Churassy, and J. B. Koffend, \href{https://doi.org/10.1103/PhysRevLett.52.267}{Phys. Rev. Lett. $\bm{52}$, 267 (1984).}

\bibitem{Tiesinga_PA_2005} E. Tiesinga, K. M. Jones, P. D. Lett, U. Volz, C. J. Williams, and P. S. Julienne, \href{https://doi.org/10.1103/PhysRevA.71.052703}{Phys. Rev. A $\bm{71}$, 052703 (2005).}

\bibitem{Tiesinga_DVR_1998} E. Tiesinga, C. J. Williams, and P. S. Julienne, \href{https://doi.org/10.1103/PhysRevA.57.4257}{Phys. Rev. A $\bm{57}$, 4257 (1998).}

\bibitem{LBE_comment} We checked the validity of our DVR method with LeRoy-Bernstein formula. The lowest potentials of $0_u$ branch with $T = 1,2$ can be approximated to from of $r^{-n}$, which follows vibrational progression. See also; R. J. LeRoy and R. B. Bernstein, \href{http://dx.doi.org/10.1063/1.1673585}{J. Chem. Phys. $\bm{52}$, 3869 (1970).}

\bibitem{Hamley_RARb_2009} C. D. Hamley, E. M. Bookjans, G. Behin-Aein, P. Ahmadi, and M. S. Chapman, \href{https://doi.org/10.1103/PhysRevA.79.023401}{Phys. Rev. A $\bm{79}$, 023401 (2009).}

\bibitem{Bohn_semianalytic_1999} John L. Bohn and P. S. Julienne, \href{https://doi.org/10.1103/PhysRevA.60.414}{Phys. Rev. A $\bm{60}$, 414 (1999).}

\bibitem{Kim_OFR_2016} M. Kim, J. Lee, J. H. Lee, Y. Shin, and J. Mun, \href{https://doi.org/10.1103/PhysRevA.94.042703}{Phys. Rev. A $\bm{94}$, 042703 (2016).}

\bibitem{Bloom_clock_2014} B. J. Bloom, T. L. Nicholson, J. R. Williams, S. L. Campbell, M. Bishof, X. Zhang, W. Zhang, S. L. Bromley, and J. Ye, \href{https://doi.org/10.1038/nature12941}{Nature $\bm{506}$, 71 (2014).} 

\bibitem{Nemitz2016} N. Nemitz,	T. Ohkubo,	M. Takamoto, I. Ushijima,	M. Das, N. Ohmae, and H. Katori, \href{https://doi:10.1038/nphoton.2016.20}{Nat. Photonics $\bm{10}$, 258 (2016).} 

\bibitem{Hofrichter2016} C. Hofrichter, L. Riegger, F. Scazza, M. Höfer, D. R. Fernandes, I. Bloch, and S. Fölling, \href{https://doi.org/10.1103/PhysRevX.6.021030}{Phys. Rev. X $\bm{6}$, 021030 (2016).}


\end{references}
\end{document}